\documentclass[aps,prl,twocolumn,amsmath,amssymb,superscriptaddress,citeautoscript,longbibliography,floatfix]{revtex4-1}

\usepackage{graphicx}
\usepackage{dcolumn}
\usepackage{bm}
\usepackage{color}
\usepackage{mathptmx}
\usepackage{mathrsfs}
\usepackage{multirow}
\usepackage{natbib}

\usepackage{epsf}
\usepackage{epsfig}
\usepackage{epstopdf}

\begin{document}

\title{Reversible adiabatic temperature change in the shape memory Heusler alloy Ni\boldmath$_{2.2}$Mn$_{0.8}$Ga: \\an effect of structural compatibility}

\author{P.\ Devi}\email{parul.devi@cpfs.mpg.de}
\affiliation{Max Planck Institute for Chemical Physics of Solids,  N\"{o}thnitzer Str.\ 40, 01187 Dresden, Germany}

\author{M.\ Ghorbani Zavareh}
\affiliation{Max Planck Institute for Chemical Physics of Solids,  N\"{o}thnitzer Str.\ 40, 01187 Dresden, Germany}

\author{C.\ Salazar Mej\'{i}a}
\affiliation{Dresden High Magnetic Field Laboratory (HLD-EMFL), Helmholtz-Zentrum Dresden-Rossendorf, 01328 Dresden, Germany}

\author {K.\ Hofmann}
\affiliation{Eduard-Zintl-Institut f\"{u}r Anorganische und Physikalische Chemie, Technische Universit\"{a}t Darmstadt, Alarich-Weiss-Str. 12, 64287 Darmstadt, Germany}

\author{B. Albert}
\affiliation{Eduard-Zintl-Institut f\"{u}r Anorganische und Physikalische Chemie, Technische Universit\"{a}t Darmstadt, Alarich-Weiss-Str. 12, 64287 Darmstadt, Germany}

\author{C.\ Felser}
\affiliation{Max Planck Institute for Chemical Physics of Solids,  N\"{o}thnitzer Str.\ 40, 01187 Dresden, Germany}

\author{M.\ Nicklas}
\affiliation{Max Planck Institute for Chemical Physics of Solids,  N\"{o}thnitzer Str.\ 40, 01187 Dresden, Germany}

\author{Sanjay Singh}\email{sanjay.singh@cpfs.mpg.de}
\affiliation{Max Planck Institute for Chemical Physics of Solids,  N\"{o}thnitzer Str.\ 40, 01187 Dresden, Germany}
\affiliation{School of Materials Science and Technology, Indian Institute of Technology (BHU), Varanasi-221005, India}

\date{\today}

\begin{abstract}
The large magnetocaloric effect (MCE) observed in Ni-Mn based shape-memory Heusler alloys put them forward to use in magnetic refrigeration technology. It is associated with a first-order magnetostructural (martensitic) phase transition. We conducted a comprehensive study of the MCE for the off-stoichiometric Heusler alloy Ni$_{2.2}$Mn$_{0.8}$Ga in the vicinity of its first-order magnetostructural phase transition. We found a reversible MCE under repeated magnetic field cycles. The reversible behavior can be attributed to the small thermal hysteresis of the martensitic phase transition. Based on the analysis of our detailed temperature dependent X-ray diffraction data, we demonstrate the geometric compatibility of the cubic austenite and tetragonal martensite phases. This finding directly relates the reversible MCE behavior to an improved geometric compatibility condition between cubic austenite and tetragonal martensite phases. The approach will help to design shape-memory Heusler alloys with a large reversible MCE taking advantage of the first-order martensitic phase transition.
\end{abstract}

\maketitle


The magnetocaloric effect (MCE) can be quantified as an isothermal magnetic entropy change ($\Delta S_{M}$) or an adiabatic temperature change ($\Delta T_{ad}$) under the application of magnetic field. It is an intrinsic magneto-thermodynamic property of magnetic materials \cite{Franco2012}. Magnetic refrigeration technology based on the MCE has higher refrigeration efficiency compared to other caloric effects \cite{Bruck2005} making it an edge over to the others. In recent past, different MCE materials have been studied and the potential candidates for magnetic refrigeration are reported as Gd$_{5}$(Si$_{1-x}$Ge$_{x}$)$_{4}$ \cite{Pecharsky1997, Moore2009, Fujita2003}, La(Fe$_{13-x}$Si$_{x}$) \cite{Hu2001, Lyubina2008}, Mn(As$_{1-x}$Sb$_{x}$) \cite{Wada2002}, MnFe(P$_{1-x}$As$_{x}$) \cite{Tegus2002}, and off-stoichiometric Heusler alloys Ni$_{2}$Mn$X$ ($X={\rm Ga}$, In, Sb and Sn) \cite{ Du2007, Krenke2005, Zavareh2015}. Among these, Ni-Mn-based Heusler alloys are the subject of special interest as they do not involve toxic and rare-earth elements, but exhibit large values of the MCE at reasonable magnetic field \cite{Krenke2005, Liu2012, Zavareh2015}.

Magnetic shape-memory Heusler alloys undergo a first-order structural phase transition from a high temperature, high symmetry cubic austenite phase to a low temperature, low symmetry martensite phase \cite{Krenke2005, Liu2012, Song2013, Zavareh2015,Singh2015, Singh2013, Singh2014}. This first-order transition leads to large $\Delta S_{M}$ and $\Delta T_{ad}$ because of both structural and magnetic contributions to the MCE \cite{Krenke2005}. For inducing a first-order phase transition energy must be spent to overcome the potential barrier between the austenite and martensite phases. This energy leads to intrinsic irreversibilities in both $\Delta S_{M}$ and $\Delta T_{ad}$, which can drastically reduce the cooling efficiency of a device. Irreversible behavior arises due to both thermal as well as magnetic hysteresis. To minimize the irreversibility, it is necessary to reduce the hysteresis \cite{Krenke2005}. Hysteresis is an inherent property of first-order phase transformation, which can be reduced by various internal parameters such as chemical composition, type and amount of a doping element as well as extrinsic parameters such as the sample preparation method, annealing conditions, applied magnetic field, pressure, heating and cooling rate, sequence of measurements and cycling \cite{Krenke2005, Du2007, Zavareh2015, Liu2012, Song2013}.  Recently, it has been proposed that the reversibility of the phase transition, i.e.\ small to no hysteresis, can be achieved by satisfying the geometric compatibility condition between austenite and martensite phases \cite{Song2013, Bhattacharya2003, Cui2006, Srivastava2010, James2000, Zhang2009, Hane1998}.

In the present work, we have studied the MCE and its relation to transformation hysteresis effects at the martensitic transition in the off-stoichiometric shape-memory Heusler alloy Ni$_{2.2}$Mn$_{0.8}$Ga. Ni$_{2.2}$Mn$_{0.8}$Ga exhibits a small hysteresis and a conventional MCE, i.e.\ the temperature increases upon application of a magnetic field. The MCE is reversible in the hysteresis region of the martensitic transformation as our $\Delta T_{ad}$ measurements in pulsed magnetic field cycles demonstrate. To investigate the origin of the reversible behavior, we conducted powder X-ray diffraction (PXRD) experiments of the martensite and the austenite phases in order to calculate and analyze the geometric transformation matrix $\mathbf{U}$. We found a geometric compatibility of both phases in Ni$_{2.2}$Mn$_{0.8}$Ga. This strongly suggests the geometric compatibility of martensite and austenite phases to be at the basis of the only small hysteresis and the reversible MCE.

\begin{figure*}[t!]
\centering
\includegraphics[width=0.9\linewidth]{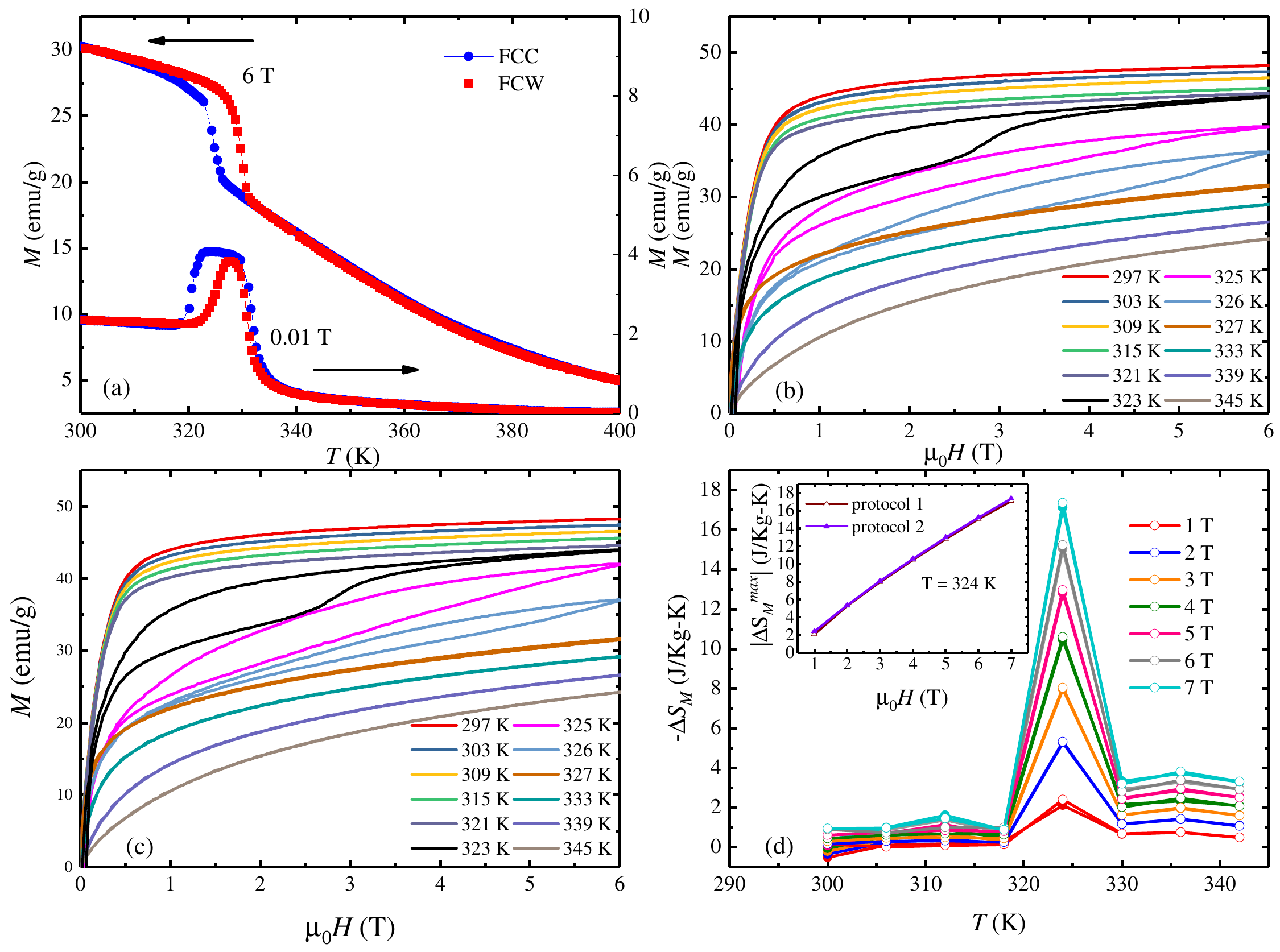}
\caption{(Color online)
(a) Field cooled cooling (FCC) and field cooled warming (FCW) magnetization $M(T)$ curves at 0.01~T and 6~T for Ni$_{2.2}$Mn$_{0.8}$Ga. (b) $M(H)$ isotherms with protocol 1 and (c) protocol 2 taken at different temperatures in static magnetic fields up to 6~T between 297 and 345~K. (d) Isothermal magnetic entropy as a function of temperature under different magnetic fields applied, solid and open circles for protocol 1 and 2, respectively. Inset: variation of  $\Delta S_{M}$   for different magnetic field at $T = 324$~K for both protocols.}
\label{magnetization}
\centering
\end{figure*}


A polycrystalline ingot with nominal composition of Ni$_{2.2}$Mn$_{0.8}$Ga  was prepared under Ar atmosphere from its pure constituent elements using an arc-melting technique. The ingot was melted six times to ensure a good homogeneity. Afterwards, the as-cast button-shaped ingot was annealed for 9 days in a sealed quartz tube under vacuum at 1100~K to obtain high homogeneity and subsequently quenched in an ice-water mixture. The compositional analysis was done by energy-dispersive spectroscopy at different spots. The average composition turns out to be Ni$_{2.19}$Mn$_{0.8}$Ga. The magnetic properties have been characterized using a magnetic properties measurement system (Quantum Design). Measurements of the MCE have been carried out in pulsed magnetic fields at the Dresden High Magnetic Field Laboratory using a home-built set up. The powder X-ray diffraction (PXRD) experiments were carried out using a StadiP diffractometer (Stoe \& Cie.)  with Mo $K\alpha$1-radiation  $\lambda$=$0.70930 {\rm~\AA}$, Ge $[1 1 1]$ monochromator.

The temperature-dependent magnetization curves $M(T)$ of Ni$_{2.2}$Mn$_{0.8}$Ga measured in external magnetic fields of 0.01~T and 6~T during cooling and heating cycles are shown in Fig.\ \ref{magnetization}. Upon cooling, the austenitic to the martensitic phase transition starts at $M_{s} = 323$~K (martensite start temperature) and ends at $M_{f}= 315$~K (martensite finish temperature). Upon heating, the reverse transformation, martensite to austenite, is found to start at $A_{s} = 320$~K (austenite start temperature) and to finish at $A_{f} = 328$~K (austenite finish temperature). The hysteresis width observed from these characteristic temperatures is about 5~K, which is small in comparison with other magnetic shape-memory Heusler alloys \cite{Khovaylo2008, Khovailo2002, Khovaylo2009, Gottschall2015, Zavareh2015, Srivastava2010}. $M(T)$ curves at 6~T shows that the magnetic fields shifts the martensitic transition toward higher temperatures. Magnetic field stabilizes the phase with the higher magnetic moment, in this case the martensitic phase. Therefore, the transition from austenite to martensite can be induced by field. However, the temperature range at which this transition can be induced is limited by the shift of the transition with field.

Motivated by the very small thermal hysteresis, we recorded the magnetization data as a function of the magnetic field using two different protocols to determine the magnetic hysteresis. Following protocol~1, the sample was heated up to 400~K to form the austenite phase, then cooled in zero field down to 200 K to ensure the complete transformation into the martensite phase, and then subsequently heated up to the measurement temperature where the $M(H)$ data were taken (see Fig.\ \ref{magnetization}b) \cite{Liu2008, Gottschall2016}. In protocol~2, the $M(H)$ loops were recorded from 297~K to 345~K one after the other without any thermal cycling as typically used for second order phase transitions (see Fig.\ \ref{magnetization}c) \cite{Pecharsky1999, Sanjay2014a}. These $M(H)$ curves show that the transition from austenite to martensite can be induced at 323, 325 and 326~K using both protocols. However, there is no significant difference in the isothermal $M(H)$ curves recorded using the two different protocols. We further calculated magnetic entropy change ($\Delta S_{M}$)  from the isothermal $M(H)$ curves using the equation \cite{Sanjay2014a}:
\begin{equation}
 \Delta\,S_{M} = S (T, H) - S (T, 0)= \int_{0}^{H} \left({\partial M\over\partial T}\right)_H  ~dH.
\end{equation}
As expected, the $\Delta S_{M}(T)$ curves also show almost identical values for both protocols (heating) at all magnetic fields (see Fig.\ \ref{magnetization}d). However, we got a minor difference between both values at 300~K in a magnetic field of 1~T. This difference in the martensite phase can stem from a twinned structure of the martensite \cite{Planes2009}.

The small thermal hysteresis, reversibility in magnetization in the region of the martensitic transformation, and similar values of $\Delta S_{M}$ obtained for both protocols indicate that Ni$_{2.2}$Mn$_{0.8}$Ga is a promising candidate for the observation of a reversible MCE and for future magnetocaloric applications.
Therefore, we investigated the MCE in details by direct measurements of $\Delta T_{ad}$ in pulsed magnetic fields. The pulsed magnetic field experiments provide the opportunity for an analysis of the temperature response of the material to magnetic field on a time scale of $\sim 1$ to 10~ms which is comparable with typical operation frequencies ($1\sim10$~Hz) of magnetocaloric cooling devices \cite{Zavareh2016}. The corresponding magnetic-field change rate is $2 - 50$~T/s, in contrast to most studies reported in literature based on steady-field experiments with typical rates of 0.01~T/s. Thus, pulsed-field studies provide a comprehensive access to the dynamics of the MCE near real operational conditions.

\begin{figure}[t!]
\includegraphics[width=0.9\linewidth]{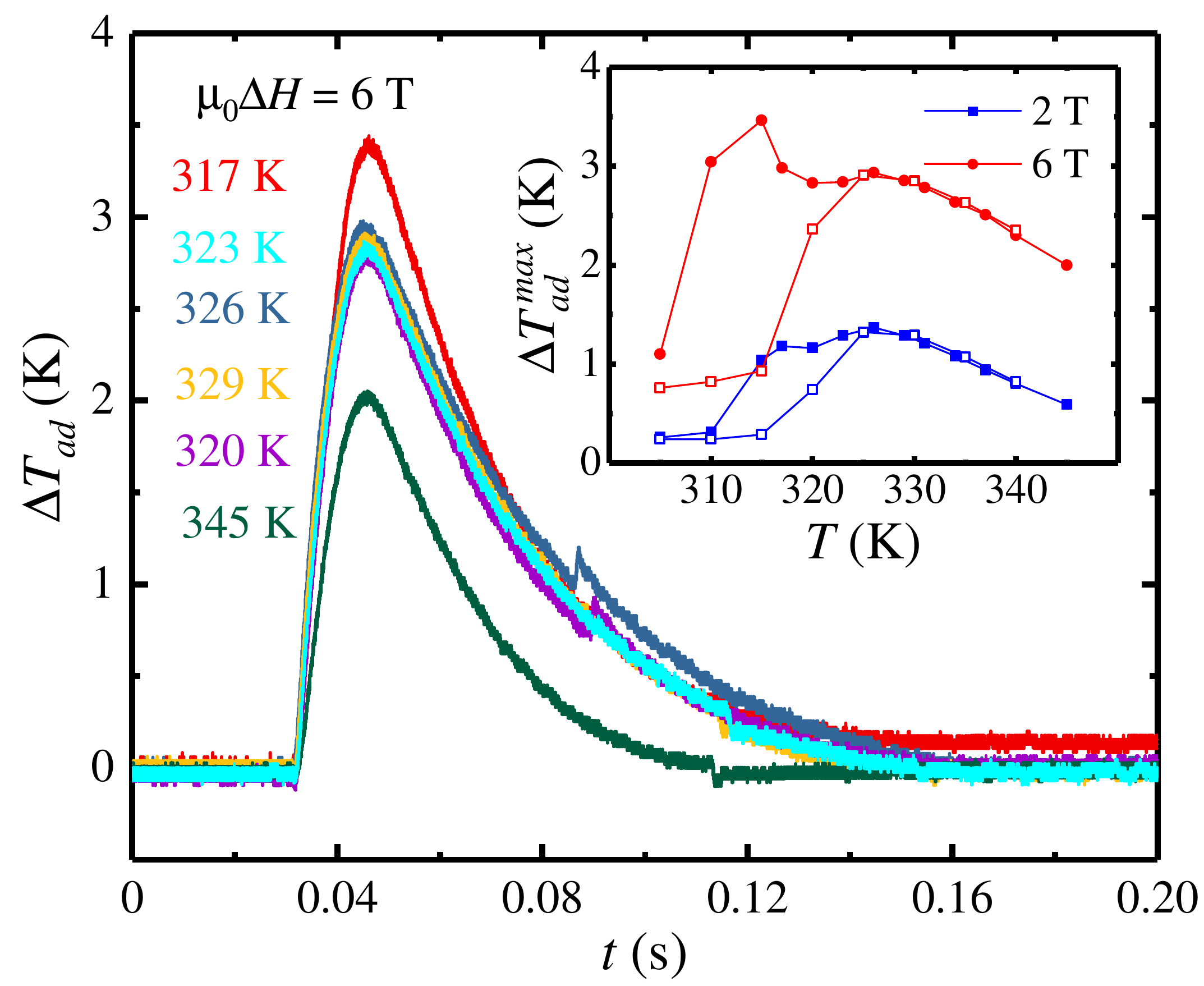}
\centering
\caption{(color online)  Adiabatic temperature change, $\Delta T_{ad}$, of Ni$_{2.2}$Mn$_{0.8}$Ga as a function of time at a magnetic field of 6~T for different temperatures during cooling. Inset: $\Delta T^{\rm max}_{ad}$  as a function of temperature for pulsed fields of 2 and 6~T. Open symbols represent data taken upon heating and closed upon cooling.}
\label{MCE}

\end{figure}

Figure \ref{MCE} shows selected $\Delta T_{ad}(t)$ curves at different temperatures for magnetic field pulses of 6~T. All plotted data were recorded after reaching the measurement temperatures during cooling from 350~K or from the previous measured temperature. Data on heating were also recorded (not shown here). $\Delta T_{ad}$ has contributions from both structural and magnetic transitions, similar to the case of Ni$_{2.19}$Mn$_{0.81}$Ga where the transitions take place in the same temperature range \cite{Pareti2003}. It is important to note that for each of the temperatures $\Delta T_{ad}(t)$ goes back to the initial value before the pulse. This indicates the reversibility of the MCE. The inset of Fig.\ \ref{MCE} displays $\Delta T^{\rm max}_{ad}$, taken at the maximum in the $\Delta T_{ad}(t)$ curve, for applied magnetic fields of 2 and 6~T, recorded both on cooling and heating. For both fields, the broad shape of the maximum in the curves of $\Delta T^{\rm max}_{ad}$, which is desirable for applications, covers a temperature window of about 35~K. Under a magnetic pulse of 6~T, $\Delta T^{\rm max}_{ad}(t)$ reaches a maximum of 3.5~K at 317~K. We note, the directly measured value of $\Delta T^{\rm max}_{ad}$ differs from that calculated from isothermal entropy change and specific heat data as expected and similar to previous studies on Ni$_{2}$MnGa and Ni$_{2.19}$Mn$_{0.81}$Ga magnetic shape memory-Heusler alloys \cite{Manosa1997, Kirkham2014, Sasso2006, Khovaylo2008}.

\begin{figure}[t!]
\includegraphics[width=0.9\linewidth]{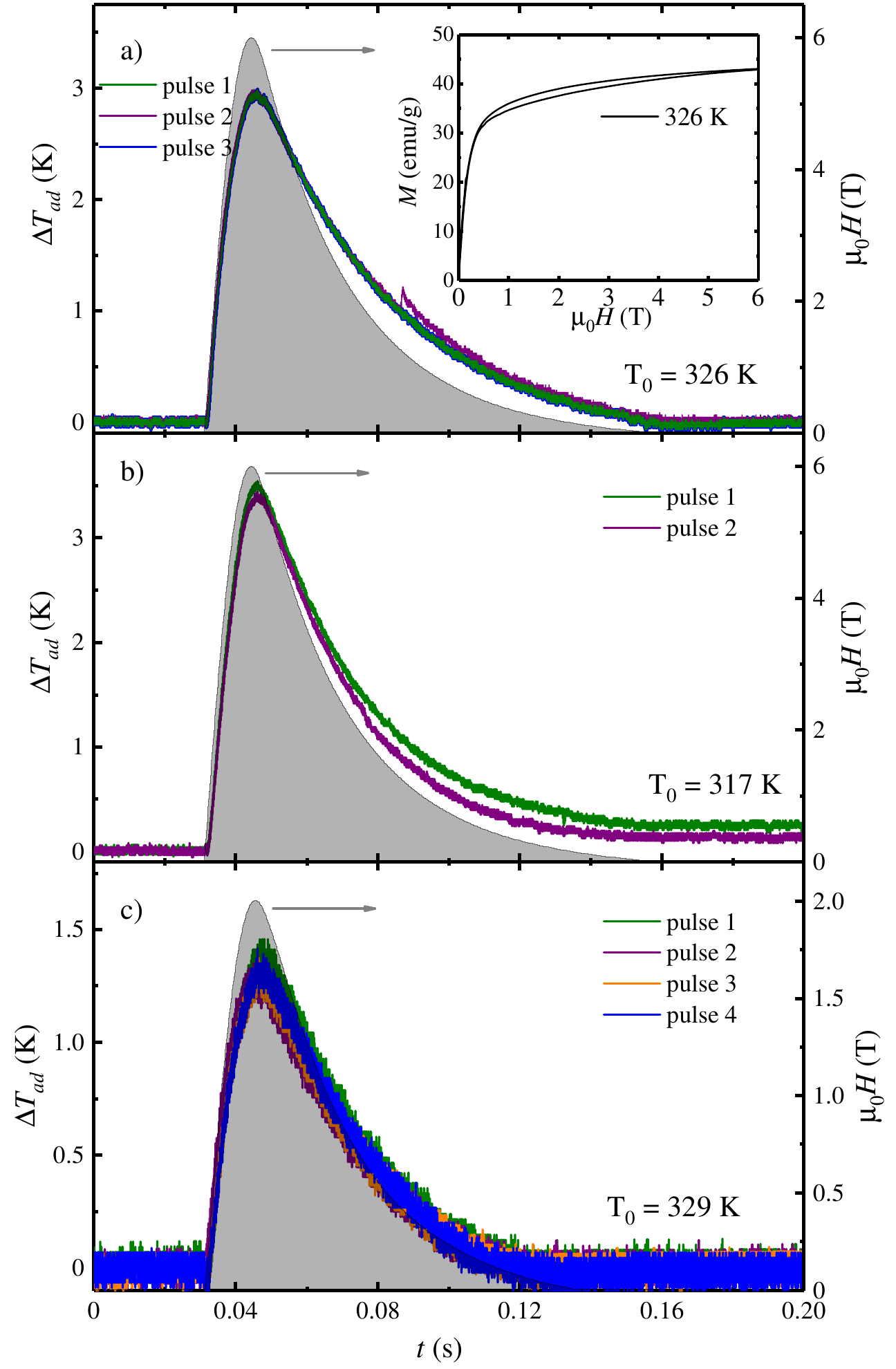}
\centering
\caption{(color online)  Time dependence of $\Delta T_{ad}$ measured at (a) 326~K and (b) 317~K reached upon cooling, for magnetic-field pulses of 6~T and (c) at 329~K reached upon heating, for magnetic-field pulses of 2~T. See text for details. The right axes refer to the magnetic field profile. Inset in (a) shows the field induced $M(H)$ curve at 326~K after cooling the sample from 400~K.}
\label{MCE_reversibility}
\end{figure}

To achieve higher efficiencies in magnetic cooling devices, the reversibility upon magnetic field cycling is crucial. To study the reversibility of the MCE in Ni$_{2.2}$Mn$_{0.8}$Ga, we have measured $\Delta T_{ad}(t)$ for three subsequent 6~T magnetic field pulses at 326~K reached upon cooling, which is just above the martensite start temperature, $M_{s} = 323$~K (see Fig.\ \ref{MCE_reversibility}a). Before pulse~1, the sample was heated to the austenite phase and subsequently cooled to the measurement temperature. Pulse~2 and 3 followed immediately after pulse~1. After pulse~1 $\Delta T_{ad}(t)$ exhibits an almost reversible behavior, only a small offset of 0.14~K remains. This value is almost unchanged for pulse~2 and pulse~3. The $M(H)$ curve at 326~K shown in the inset of Fig.\ \ref{MCE_reversibility}a clearly indicates that the field induced transition from austenite to the martensite. We repeated the previously described experiment after further cooling down to 317~K. This temperature is in between the martensite start ($M_{s} = 323$~K)  and martensite finish ($M_{f} = 315$~K) temperatures (see  Fig.\ \ref{MCE_reversibility}b). After pulse 1  we found only a tiny irreversible offset of 0.26~K. After pulse~2 the offset of 0.13~K was even smaller, while the values of $\Delta T^{\rm max}_{ad}$ for both pulses were almost the same. We note that the recorded offsets are smaller than the uncertainty in the measurement of $\Delta T$ \cite{Zavareh2016}. Additionally, we investigated the irreversibility of the MCE at 329~K ($T>A_s$). Here, the measurement temperature was approached upon heating from well below the martensitic transition. At 329~K four consecutive magnetic pulses up to 2~T were applied. As can be seen in Fig.\ \ref{MCE_reversibility}c, $\Delta T_{ad}(t)$ is reversible for all pulses.
Thus, the previous results are a fair indication of the fast kinetics of the thermoelastic transformation which is reversible due to the small hysteresis. Moreover, the pulsed magnetic fields measurements give evidence that Ni$_{2.2}$Mn$_{0.8}$Ga exhibits an almost perfect reversible MCE on the time scale of magnetocaloric devices.


In shape-memory Heusler alloys the occurrence of hysteresis, and consequently, an irreversible behavior of the MCE at the martensitic transformation, is closely related to the austenite and martensite phases and their interfaces. In most cases, this interface is a plane, known as habit plane. During the phase transformation from austenite to martensite an elastic transition layer forms at the interface instead of an exact interface between both phases. For forward and reverse transformation, the energy associated with the formation of an austenite/martensite interfaces results in hysteresis \cite{Srivastava2010, Zarnetta2010}. Recently, it has been shown that this hysteresis can be overcome by improving the compatibility condition between austenite and martensite phases \cite{Bhattacharya2003, Song2013, Cui2006, Srivastava2010}. The information about the compatibility of both phases is contained in the deformation matrix, which is calculated from the lattice parameters of both phases \citep{Bhattacharya2003}.

The martensitic transformation is diffusionless. The lattice vectors of both austenite and martensite phases are related by a homogenous $3\times3$ deformation matrix $\mathbf{U}$. This matrix $\mathbf{U}$ is called Bain distortion matrix or the transformation matrix. The determinant of this matrix $\mathbf{U}$ represents the volume change between the two phases. For the martensite and austenite phase to be compatible or for the formation of an exact interface between austenite and martensite the determinant of the transformation matrix $\mathbf{U}$ should be one. This is called geometric compatibility condition for the material going from the cubic austenite to the martensite phase \cite{Bhattacharya2003}. The transformation matrix $\mathbf{U}$ and the number of modifications of martensite (tetragonal, monoclinic and orthorhombic) vary for different systems \cite{Bhattacharya2003, Cui2006, Song2013}.

\begin{figure}[t!]
\centering
\includegraphics[width=0.9\linewidth]{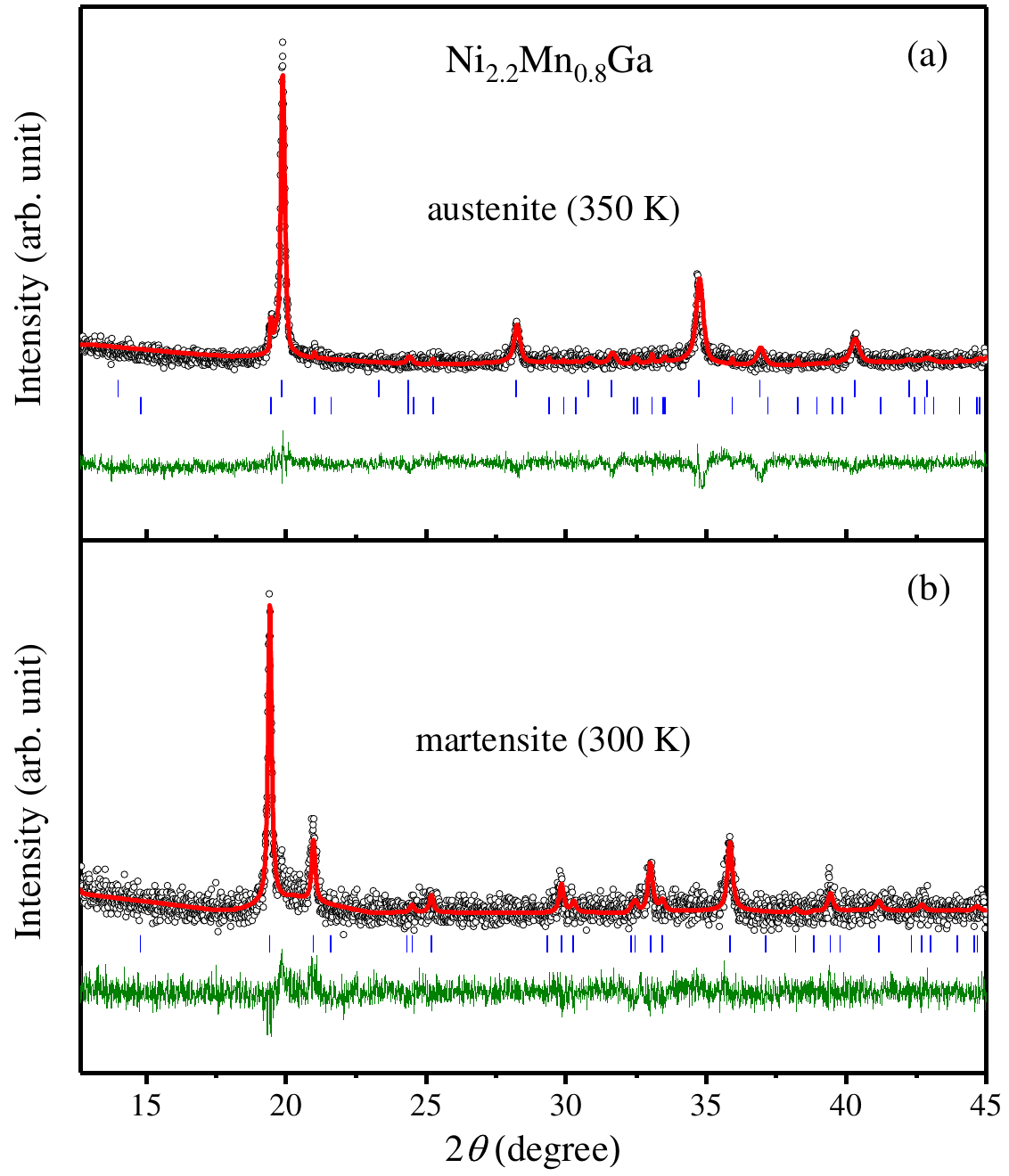}
\caption{(Color online) Lebail refinements for PXRD patterns of Ni$_{2.2}$Mn$_{0.8}$Ga in (a) austenite and (b) martensite phases. The experimental data are shown by circles and the fitted curve and residue by lines, respectively. The ticks represent Bragg-peak positions.}
\label{XRD}
\end{figure}

To determine the transformation matrix $\mathbf{U}$ for Ni$_{2.2}$Mn$_{0.8}$Ga the structure information for both phases is needed. PXRD experiments were conducted at 350~K and 300~K, to obtain data for the austenite and martensite phase. The LeBail fits of the PXRD patterns of both phases are shown in Fig.\ \ref{XRD}.  At room temperature, Ni$_{2.2}$Mn$_{0.8}$Ga is in martensitic phase ($M_{s}= 323$~K, see also Fig.\ \ref{magnetization}). All of the reflections in the PXRD pattern could be indexed based on a body-centered tetragonal lattice (space group $I4$/$mmm$) and the lattice parameters were refined to $a=3.9013(6){\rm~\AA}$ and $c=6.5129(4){\rm~\AA}$, while at 350~K Ni$_{2.2}$Mn$_{0.8}$Ga is in the austenitic phase and has a cubic structure (space group $Fm$-3$m$). The refined lattice parameter is $a=5.8286(2){\rm~\AA}$. A small fraction of  the martensite phase coexists at 350~K, which can be attributed to the effect of a residual stress generated upon grinding the ingot into powder \cite{Singh2015,  Singh2013, Singh2014}.

In general, the cubic to tetragonal transformation can be described by two unequal stretches. The number of possible variants of martensite are determined by the number of rotations that are possible in the point group of the austenite $P_{a}$ divided by the number of rotations that are possible in the point group of the martensite $P_{m}$. The number of rotations possible in the point group of the cubic austenite is 24, whereas the number of rotations possible in the point group of the martensite is 8. So, the cubic to tetragonal transformation results in three variants of martensite which are of course related by symmetry and must have the same eigenvalue \cite{Hane1998}. The variants of martensite for face-centered cubic to face-centered tetragonal described as follows \cite{Bhattacharya2003}:
\begin{equation}
\mathbf{U}_{1}=
\begin{pmatrix}
\beta & 0 & 0\\
0 & \alpha & 0\\
0 & 0 & \alpha
\end{pmatrix}
\mathrm{,~}
\mathbf{U}_{2}=
\begin{pmatrix}
\alpha & 0 & 0\\
0 & \beta & 0\\
0 & 0 & \alpha
\end{pmatrix}
\mathrm{and~}
\mathbf{U}_{3}=
\begin{pmatrix}
\alpha & 0 & 0\\
0 & \alpha & 0\\
0 & 0 & \beta
\end{pmatrix}\mathrm{.}
\label{U}
\end{equation}
The transformational stretches $\alpha$ and $\beta$ are derived from the lattice parameters of cubic austenite and face centered tetragonal martensite phases are $\alpha=\frac{a_{F}}{a_0}$ and $\beta=\frac{c_{F}}{a_0}$, where the index $F$ stands for face centered. In our case, for a transformation from the face-centered cubic ($Fm$-3$m$) to the body-centered tetragonal ($I4$/$mmm$) structure the lattice parameters of body-centered unit cell can be converted to the face-centered unit cell by the following relationships: $a=a_{F}=\sqrt{2}a_{I}$ and $c=c_{F}=c_{I}$ \cite{Banik2007}. Here, the index $I$ stands for body centered. These stretches satisfy $\alpha > 0$, $\beta > 0$ and $\alpha \neq \beta$ \cite{Hane1998}.

Thus, the transformation matrix of one of the corresponding martensite variants of Ni$_{2.2}$Mn$_{0.8}$Ga is:
\begin{equation}
\mathbf{U}_{1}=
\begin{pmatrix}
1.1174 & 0 & 0\\
0 & 0.9466 & 0\\
0 & 0 & 0.9466
\end{pmatrix}.
\end{equation}
$\mathbf{U}_2$ and $\mathbf{U}_3$ follow directly from $\mathbf{U}_1$ according to Eq.\ \ref{U}.
The determinant of this transformation matrix is very close to one $|\mathbf{U}|=1.0012$. The deviation from unity is only 0.12\%  which is substantially smaller in comparison to the previous studies \cite{Liu2005, Banik2007}. Mn$_{2}$NiGa exhibits a thermal hysteresis of 50~K \cite{Liu2005}. From the lattice parameters one obtains $|\mathbf{U}|=0.9936$ with a variation of  0.64\% from unity \cite{Liu2005}. A slightly different composition, Ni$_{2.2}$Mn$_{0.75}$Ga, from our studied material Ni$_{2.2}$Mn$_{0.8}$Ga exhibits a hysteresis of 14~K in $|\mathbf{U}|=0.9939$ which differs 0.61\% from unity \cite{Banik2007}. These values deviate significantly more in comparison to our study. Hence, exemplifying the validity of the geometric compatibility condition of the austenite and martensite phases in Ni$_{2.2}$Mn$_{0.8}$Ga.

In general, in shape-memory Heusler alloys the directly measured $\Delta T_{ad}(t)$ is expected to be influenced by the width of the hysteresis as well as the sharpness of the martensitic transition. Another factor that can also affect the reversibility of $\Delta T_{ad}(t)$ is a kinetic arrest due to a structurally and magnetically inhomogeneous state. In case of an alloy with a reduced hysteresis, the lattice coherence results in faster kinetics of the magnetostructural transformation and, thus, in a smaller energy barrier at the interface, consistent with our results.


In summary, we have studied the reversible adiabatic temperature change in the shape-memory Heusler alloy Ni$_{2.2}$Mn$_{0.8}$Ga and its relation to the structural properties at the martensitic transformation. We found that the reversibility of MCE is directly related to the small thermal and magnetic hysteresis which is based on the geometric compatibility of the austenite and martensite phases in Ni$_{2.2}$Mn$_{0.8}$Ga. Therefore, we can attribute the reversible behavior to the highly mobile transition layer between the two phases that leads to a reduction of the energy required for creating interfaces. Our finding provides a pathway to improve the reversibility of the MCE in shape-memory Heusler alloys in the region of their martensitic transformation based on the geometric compatibility of the austenite and martensite phases.

\begin{acknowledgments}
This work was financially supported by ERC Advanced Grant No.\ 291472 `Idea Heusler'.  We acknowledge the support of the HLD at HZDR, member of the European Magnetic Field Laboratory (EMFL). S.S. thanks Alexander von Humboldt foundation, Germany and Science and Engineering Research Board of India for financial support through the award of Ramanujan Fellowship.
\end{acknowledgments}

\bibliography{MCE_NiMnGa}

\end{document}